\documentclass[twocolumn,showpacs,preprintnumbers,amsmath,amssymb,prl]{revtex4}
\usepackage{dcolumn}% Align table columns on decimal point
\usepackage{bm}% bold math

\usepackage[dvips]{graphicx}
\usepackage{wrapfig,subfigure}
\usepackage{amssymb}
\usepackage{color}

\begin{document}

\title{Quantum interference in the classically forbidden region: A parametric
oscillator}

\author{ M. Marthaler$^{1,2}$ and M. I. Dykman$^1$}
 \affiliation{
$^1$Department of Physics and~Astronomy,~Michigan State
 University,~East~Lansing,~MI 48824,~USA\\
 $^2$Institut f\"ur Theoretische Festk\"orperphysik and DFG-Center
for Functional Nanostructures (CFN), Universit\"at Karlsruhe,
D-76128 Karlsruhe, Germany}
\date{\today}

\begin{abstract}
We study tunneling between period-2 states of a parametrically
modulated oscillator. The tunneling matrix element is shown to
oscillate with the varying frequency of the modulating field. The
effect is due to spatial oscillations of the wave function and the
related interference in the classically forbidden region. The
oscillations emerge already in the ground state of the oscillator
Hamiltonian in the rotating frame.
\end{abstract}

\pacs{74.50.+r, 03.65.Xp,  05.45.-a,  05.60.Gg}
\maketitle

Nonlinear micro- and mesoscopic vibrational systems have attracted
much interest in recent years. In such systems damping is often
weak, and even a comparatively small resonant field can lead to
bistability, i.e., to coexistence of forced vibrations with
different phases and/or amplitudes. Quantum and classical
fluctuations cause transitions between coexisting vibrational
states. The transitions are not described by the conventional theory
of metastable decay, because the states are periodic in time and the
systems lack detailed balance. Experimentally, classical transition
rates have been studied for such diverse vibrational systems as
modulated trapped electrons \cite{Lapidus1999}, Josephson junctions
\cite{Siddiqi2005}, nano- and micromechanical oscillators
\cite{Aldridge2005,Stambaugh2006b,Almog2007}, and trapped atoms
\cite{Kim2005}, and the results are in agreement with theory
\cite{Dykman1979a,Dykman1998}.

Currently much experimental effort is being put into reaching the
quantum regime \cite{Schwab2005a,Boaknin2007}. In this regime
tunneling between coexisting classically stable periodic states
should become important, for weak dissipation. It was first studied
for a resonantly driven oscillator, where a semiclassical analysis
\cite{Sazonov1976} made it possible to find the tunneling exponent
in a broad parameter range \cite{Dmitriev1986a}.

Tunneling is particularly interesting for a parametrically modulated
oscillator. Here, the coexisting classical periodic states have
period $2\tau_F$, where $\tau_F$ is the modulation period. Such
period-2 states are identical except that the vibrations are shifted
in phase by $\pi$. Therefore the corresponding quantum states
(Floquet states) are degenerate. Tunneling should lift this
degeneracy, as for a particle in a symmetric static double-well
potential. Earlier the tunneling matrix element was found
\cite{Wielinga1993} for modulation at exactly twice the oscillator
eigenfrequency $\omega_0$. Recently the tunneling exponent was
obtained in a general case where the modulation frequency
$\omega_F=2\pi/\tau_F$ is close to $2\omega_0$ \cite{Marthaler2006}

In this paper we show that tunneling between period-2 states of a
parametrically modulated oscillator displays unexpected features. We
find that the tunneling matrix element oscillates with varying
$\omega_F-2\omega_0$, periodically passing through zero. These
oscillations are accompanied by, and are due to spatial oscillations
of the wave function in the classically forbidden region.
\begin{figure}[h]
\begin{center}
\includegraphics[width=2.4in]{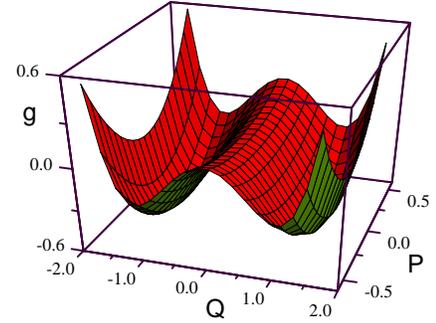}
\caption{(Color online) The scaled effective Hamiltonian of the
oscillator in the rotating frame $g(Q,P)$, Eq.~(\ref{eq:g}), for
$\mu=0.5$. The minima of $g(Q,P)$ correspond to the period-2
vibrations. The eigenvalues of $\hat g$ give scaled oscillator
quasienergies.}\label{fig:g}
\end{center}
\end{figure}

For resonant modulation, $|\omega_F-2\omega_0| \ll \omega_F$, and
for a small amplitude of the modulating field $F$ the oscillator
dynamics is well described by the rotating wave approximation (RWA)
\cite{LL_Mechanics2004}. The scaled RWA Hamiltonian $\hat g$ as a
function of the oscillator coordinate $Q$ and momentum $P$ in the
rotating frame is independent of time. In a broad parameter range it
has a symmetric double-well form shown in Fig.~\ref{fig:g}. The
minima correspond to the classical period-2 states, in the presence
of weak dissipation. Respectively, of utmost interest are tunneling
transitions between the lowest single-well quantum states of $\hat
g$.

A simple model of a nonlinear oscillator that describes many
experimental systems, cf.
Refs.~\onlinecite{Lapidus1999,Siddiqi2005,Aldridge2005,Stambaugh2006b,Almog2007,Kim2005},
is a Duffing oscillator. The Hamiltonian of a parametrically
modulated Duffing oscillator has the form
\begin{equation}
\label{eq:H_0(t)}
H_0=\frac{1}{2}p^2+\frac{1}{2}\left(\omega_0^2+F\cos\omega_F
t\right)q^2+\frac{1}{4}\gamma q^4\, .
\end{equation}
For $\omega_F$ close to $2\omega_0$ and for comparatively small $F$,
\begin{eqnarray}
\label{eq:delta_omega}
\delta\omega=\frac{1}{2}\omega_F-\omega_0,\qquad
|\delta\omega|\ll\omega_0,  \qquad F\ll\omega_0^2,
\end{eqnarray}
even where the oscillator becomes bistable its nonlinearity remains
relatively small,  $|\gamma\langle q^2 \rangle|\ll\omega_0^2$. For
concreteness we set $\gamma>0$; the results for $\gamma<0$ can be
obtained by replacing $\delta\omega\to -\delta\omega$ in the final
expressions.

To describe a weakly nonlinear oscillator it is convenient to make a
canonical transformation from $q$ and $p$ to the slowly varying
coordinate $Q$ and momentum $P$,
\begin{eqnarray}
\label{eq:transformation}
 U^{\dagger}qU&=&C_{\rm
par}\left[P\cos(\omega_F t/2)-Q\sin(\omega_F
t/2)\right],\\
U^{\dagger}pU&=&-C_{\rm par}\omega_F\left[P\sin(\omega_F
t/2)+Q\cos(\omega_F
 t/2)\right]/2,\nonumber
\end{eqnarray}
where $C_{\rm par}=(2F/3\gamma)^{1/2}$ and
\begin{equation}\label{eq:L}
[P,Q]=-i\lambda\, ,\qquad \lambda=3\gamma\hbar/F\omega_F\, .
\end{equation}
The dimensionless parameter $\lambda$ plays the role of $\hbar$ in
quantum dynamics in the rotating frame \cite{Marthaler2006}.

The transformed oscillator Hamiltonian has the form
$(F^2/6\gamma)\,\hat{g}$, where $\hat g\equiv g(Q,P)$,
\begin{eqnarray}
 \label{eq:g}
g(Q,P) =
\frac{1}{4}\left(P^2+Q^2\right)^2+\frac{1}{2}(1-\mu)P^2-\frac{1}{2}(1+\mu)Q^2\,
\end{eqnarray}
[we use here a more conventional notation $g(Q,P)$ instead of
$g(P,Q)$ used in Ref.~\onlinecite{Marthaler2006}]. The terms
$\propto\exp(\pm in\omega_Ft)$ with $n\geq 1$ in $\hat g$ have been
disregarded.

The time-independent operator $\hat g$ is the scaled oscillator
Hamiltonian in the rotating frame. Its eigenvalues multiplied by
$F^2/6\gamma$ give oscillator quasienergies, or Floquet eigenvalues.
Formally, $\hat g$ is a Hamiltonian of an auxiliary stationary
system with variables $Q,P$, and the eigenvalues of $\hat g$ give
the energies of this system. The operator $\hat g$ depends on one
parameter
\begin{equation}
\label{eq:mu}
\mu=2\omega_F\delta\omega/F\, .
\end{equation}
For $\mu>-1$, $g(Q,P)$ has two minima located at $P=0$,
$Q=\pm(\mu+1)^{1/2}$. For $\mu\leq 1$ the minima are separated by a
saddle at $P=Q=0$, as shown in Fig.~\ref{fig:g}. When friction is
taken into account, the minima become stable states of period-2
vibrations. The function $g(Q,P)$ is symmetric as a consequence of
the time translation symmetry: the change $(P,Q)\rightarrow(-P,-Q)$
corresponds to shifting time in Eq.~(\ref{eq:transformation}) by the
modulation period $\tau_F$.

We assume the effective Planck constant $\lambda$ to be the small
parameter of the theory, $\lambda \ll 1$. Then the low-lying
eigenvalues of $\hat g$ form doublets. Splitting of the doublets is
due to tunneling between the wells of $g(Q,P)$. Since
$g(Q,P)=g(-Q,-P)$ is symmetric, the problem of level splitting seems
to be similar to the standard problem of level splitting in a
double-well potential \cite{LL_QM81}. As in this latter case, we
will analyze it in the WKB approximation.

The major distinction of the present problem comes from the
difference between the structure of $g(Q,P)$ and the Hamiltonian
considered in Ref.~\onlinecite{LL_QM81}. The momentum $P(Q;g)$ as
given by equation $g(Q,P)=g$ has 4 branches, with both real and
imaginary parts in the classically forbidden region of $Q$. This
leads to new features of tunneling and requires a modification of
the method \cite{LL_QM81}.

We will consider splitting $\delta g$ of the two lowest eigenvalues
of $\hat g$. Because of the symmetry, the corresponding wave
functions $\psi_{\pm}(Q)$ are
\begin{equation}
\label{eq:psi_pm}
 \psi_{\pm}(Q)=\frac{1}{\sqrt{2}}\left[\psi_l(Q)\pm
\psi_l(-Q)\right],
\end{equation}
where $\psi_l(Q)$ is the ``single-well" wave function of the left
well of $g(Q,P)$ in Fig.~\ref{fig:g}. It is maximal at the bottom of
the well $Q_{l0}=-(1+\mu)^{1/2}$ and decays away from the well. To
the leading order in $\lambda$, the corresponding lowest eigenvalue
of $\hat g$ is $g_{\min}+g_q$, where $g_{\min}=-(1+\mu)^2/4$ is the
minimum of $g(Q,P)$ and $g_q=\lambda(\mu+1)^{1/2}$ is the quantum
correction.

The wave function $\psi_l(Q)$ is particularly simple for $\mu < 0$.
In the classically forbidden region between the wells,
$|Q|<|Q_{l0}|$, it has the form
\begin{eqnarray}
\label{eq:tail}
 && \psi_l =C\left[-i\partial_P
g\right]^{-1/2}\exp[iS_0(Q)/\lambda],
\end{eqnarray}
where $S_0(Q)$ is given by the equation
$g(Q,\partial_QS_0)=g_{\min}+g_q$,
\begin{eqnarray}
\label{eq:P_pm}
 S_0(Q)&=&\int\nolimits_{Q_{l0}+L_q}^Q\,P_-(Q')\,dQ', \\
 P_{\pm}(Q)&=&i\left[1 +Q^2-\mu \pm
2\left(Q^2-\tilde\mu\right)^{1/2}\right]^{1/2},\nonumber\\
 \tilde\mu =\mu-g_q,&& \quad L_q
=\lambda/g_q^{1/2}\equiv \lambda^{1/2}(\mu + 1)^{-1/4}.\nonumber
\end{eqnarray}
We keep here only the contribution from the branch $P_-(Q)$, because
$P_-(Q)$ is zero on the boundary of the classically forbidden range
$Q_{l0}+L_q$. For $-\mu\gg \lambda$ and $|Q_{l0}+L_q|>|Q|$ the
action $S_0(Q)$ is purely imaginary. The wave function $\psi_l(Q)$
monotonically decays with increasing $Q$.

The prefactor in the wave function (\ref{eq:tail}) is determined by
the complex classical speed of the oscillator
\begin{equation}
\label{eq:prefactor_general}
 \partial_Pg = 2P_-(Q)\left(Q^2-\tilde\mu\right)^{1/2}.
\end{equation}
The normalization constant $C$ in Eq.~(\ref{eq:tail}),
\begin{equation}
\label{eq:normalization} C=[(\mu + 1)/\pi]^{1/4}\exp(-1/4),
\end{equation}
is obtained by matching, in the range $L_q\ll Q-Q_{l0}\ll |Q_{l0}|$,
Eq.~(\ref{eq:tail}) to the tail of the Gaussian peak of
$\psi_l(Q)$, which is centered at $Q_{l0}$.

We are most interested in the parameter range $\mu \gg \lambda$
where tunneling displays unusual behavior. For such $\mu$ the
momentum $P_-(Q)$ becomes complex in the range $|Q|<\tilde\mu$. This
means that the decay of the wave function is accompanied by
oscillations. To correctly describe them we had to keep corrections
$\propto g_q$ in Eq.~(\ref{eq:P_pm}).

We first rewrite Eq.~(\ref{eq:P_pm}) in the form
\begin{eqnarray}
\label{eq:asymptotic_P-}
 P_-(Q)\approx i\left[1-(Q^2-\tilde\mu)^{1/2}
 -\frac{g_q/2}{1-(Q^2-\tilde\mu)^{1/2}}\right].
\end{eqnarray}
Eq.~(\ref{eq:asymptotic_P-}) applies for $Q-Q_{l0}\gg L_q$. It is
seen that $P_-(Q)$ has two branching points inside the classically
forbidden region. The closest to $Q_{l0}$ is the point
$Q_{br}=-\tilde\mu^{1/2}$. The WKB approximation breaks down for
small $Q+\tilde\mu^{1/2}$. The wave function in this region can be
shown to be proportional to Airy function ${\rm
Ai}[-(Q+\tilde\mu^{1/2})(2\tilde\mu^{1/2}/\lambda^2)^{1/3}]$.
Therefore $\psi_l$ oscillates with $Q$ for positive
$Q+\tilde\mu^{1/2}$.

In contrast to the standard WKB theory of the
turning point, the prefactor in $\psi_l$ contains two factors that
experience branching at $-\tilde\mu^{1/2}$, see
Eqs.~(\ref{eq:tail}), (\ref{eq:prefactor_general}). The full
solution in the oscillation region can be obtained by going around
$-\tilde\mu^{1/2}$ in the complex plane following the prescription
\cite{LL_QM81}. For $\lambda^{2/3}\ll Q+\tilde\mu^{1/2}$ it gives
\begin{eqnarray}
\label{eq:oscillating}
 &&\psi_l\approx 2C|\partial_P
g|^{-1/2}\exp \left[-{\rm Im}\,S_0(Q)/\lambda\right]\cos\Phi(Q),
\nonumber\\
&&\Phi(Q)=\Phi_1(Q) + \Phi_2(Q).
\end{eqnarray}

The term Im~$S_0(Q)$ in the amplitude of the wave function
(\ref{eq:oscillating}) is determined by Eq.~(\ref{eq:P_pm}). The
phase $\Phi(Q)$ has two terms. The term $\Phi_1(Q)$ comes from the
exponential factor in the WKB wave function (\ref{eq:tail}),
\begin{equation}
\label{eq:Phi_1}
 \Phi_1(Q)=\lambda^{-1}
\int\nolimits_{-\tilde\mu^{1/2}}^Q {\rm Re}\,P_-(Q)\,dQ,
\end{equation}
where Re~$P_-(Q)$ is given by Eq.~(\ref{eq:asymptotic_P-}) in which
we set $(Q^2-\tilde\mu)^{1/2}\to i(\tilde\mu-Q^2)^{1/2}$; therefore
Re~$P_-(Q)>0$. It is simple to write $\Phi_1$ and Im~$S_0(Q)$ in
explicit form.

The term $\Phi_2(Q)$ in Eq.~(\ref{eq:oscillating}) comes from the
prefactor in $\psi_l(Q)$, Eq.~(\ref{eq:tail}),
\begin{equation}
\label{eq:prefactor_induced_phase}
 \Phi_2(Q)\approx \frac{1}{2}\arcsin\left(\frac{\mu-Q^2}{1+\mu
 - Q^2}\right)^{1/2}-\frac{\pi}{4}.
\end{equation}

\begin{figure}[ht]
\begin{center}
\includegraphics[width=2.8in]{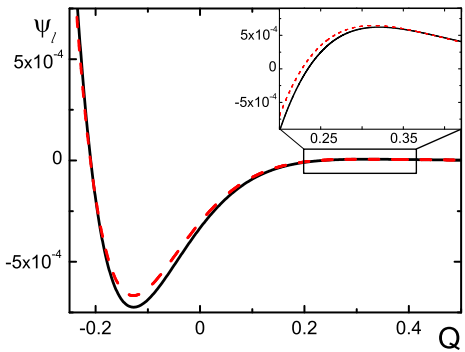}
\caption{(Color online) The wave function of the ground state in the
left well $\psi_l(Q)$ in the oscillation region for $\lambda=0.09$
and $\mu=0.5$. The solid line shows explicit expressions
(\ref{eq:oscillating})-(\ref{eq:prefactor_induced_phase}), the
dashed line shows numerical results. Inset: $\psi_l(Q)$ near its
second zero with higher resolution.}\label{fig:wavefunction}
\end{center}
\end{figure}

Decay and oscillations of the wave function described by
Eq.~(\ref{eq:oscillating}) are compared in
Fig.~\ref{fig:wavefunction} with the results of a numerical solution
of the Schr\"odinger equation $\hat g\psi= g\psi$. The left-well
wave function was obtained numerically as a sum of the two
lowest-eigenvalues solutions, cf. Eq.~(\ref{eq:psi_pm}). In this
calculation the basis of 120 oscillator Fock states was used. A good
agreement between analytical and numerical results is seen already
for not too small $\lambda =0.09$.

The above solution allows us to find the tunnel splitting $\delta
g=g_- - g_+$ of the symmetric and antisymmetric states
(\ref{eq:psi_pm}). Following the standard approach for a symmetric
double-well potential \cite{LL_QM81} we multiply the Schr\"odinger
equations for the involved states $\hat g\psi_l= g_l\psi_l$ and
$\hat g\psi_{\pm}= g_{\pm}\psi_{\pm}$ by $\psi^*_{\pm}$ and
$\psi^*_l$, respectively, integrate over $Q$ from $-\infty$ to 0 and
subtract the results. This gives
\begin{eqnarray*}
\delta g= -\lambda^2\left\{2(1-\mu)\psi_l(0)\psi^{\prime}_l(0)
\right.\nonumber\\ -
\left.\lambda^2\left[\psi_l(0)\psi^{\prime\prime\prime}_l(0) +
\psi^{\prime}_l(0)\psi^{\prime\prime}_l(0)\right]\right\}
\end{eqnarray*}
or, with account taken of Eq.~(\ref{eq:oscillating}),
\begin{eqnarray}
\label{eq:splitting_explicit}
 \delta g &=& \frac{16\lambda^{1/2}(\mu+1)^{5/4}}{(\pi\mu)^{1/2}}e^{-A/\lambda}
 \cos \left[2\Phi_1(0)\right],\\
 A&=&(\mu
 +1)^{1/2}+\mu\ln\left(\mu^{-1/2}\left[1+(\mu+1)^{1/2}\right]\right),\nonumber\\
 &&2\Phi_1(0)=\pi(\mu\lambda^{-1}-1)/2 \qquad (\mu\gg \lambda).\nonumber
\end{eqnarray}
Clearly, $\delta g$ may be positive or negative, that is, the
symmetric state may have a lower or higher quasienergy than the
antisymmetric state.

\begin{figure}[h]
\begin{center}
\includegraphics[width=2.7in]{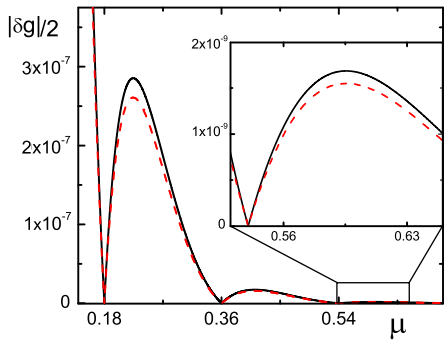}
\caption{(Color online) Scaled matrix element of tunneling between
period-2 states as a function of the scaled detuning of the
modulation frequency from twice the oscillator eigenfrequency. The
solid lines show explicit expression (\ref{eq:splitting_explicit}),
the dashed lines show the result of numerical calculations. Inset: a
higher-resolution plot of $|\delta g|/2$ vs. $\mu$ near the zero of
$\delta g$ at $\mu=6\lambda$. The data refer to $\lambda
=0.09$.}\label{fig:tunnel_splitting}
\end{center}
\end{figure}

The dimensional splitting $(F^2/6\gamma)|\delta g|$ gives twice the
matrix element of tunneling between period-2 states of the
oscillator. This matrix element has an exponential factor
$\exp(-A/\lambda)$ \cite{Marthaler2006}. In addition, it contains a
factor oscillating as a function of the scaled frequency detuning
$\mu/\lambda = 6\omega_F^2(\omega_F-2\omega_0)/3\gamma\hbar$. The
oscillation period is $\Delta(\mu/\lambda)=4$. These oscillations
are shown in Fig.~\ref{fig:tunnel_splitting}.

The oscillations of $\delta g$ result from the wave function
oscillations in the classically forbidden region. This can be seen
from the analysis of $\psi_l(Q)$ near the positive-$Q$ boundary of
the oscillation region, $Q=\tilde\mu^{1/2}$. The wave function for
$Q-\tilde\mu^{1/2}\gg \lambda$ is a combination of the WKB waves
with imaginary momenta $P_{\pm}(Q) \approx i[1\pm
(Q^2-\tilde\mu)^{1/2}]$. The coefficients in this combination can be
found in a standard way \cite{LL_QM81}. They are determined by the
phase $\Phi(\tilde\mu^{1/2})$. Only the wave with $P_-(Q)$
contributes to the tunneling amplitude, since $P_+$ remains
imaginary in the right well of $g(Q,P)$. For
$\Phi(\tilde\mu^{1/2})=(4n-3)\pi/4$ this wave has zero amplitude,
leading to $\delta g=0$. By noting that $\Phi(\tilde\mu^{1/2})=
2\Phi_1(0)-\pi/4$, we immediately obtain from
Eq.~(\ref{eq:splitting_explicit}) that $\delta g=0$ for
$\mu=2n\lambda$ with integer $n$, in agreement with
Fig.~\ref{fig:tunnel_splitting}

The occurrence of spatial oscillations of the ground state wave
function of the scaled Hamiltonian $\hat g$ does not contradict the
oscillation theorem, because $\hat g$ is not a sum of the kinetic
and potential energies and is quartic in $P$. The motion described
by the Hamiltonian $g(Q,P)$ is classically integrable. Respectively,
the quantum problem is different from dynamical tunneling in
classically chaotic systems
\cite{Tomsovic1994,Hensinger2001,Steck2001}; the effect we discuss
has not been considered for such systems, to the best of our
knowledge.

The effect is also qualitatively different from
photon-assisted/suppressed tunneling in systems with stationary
double-well potentials: our oscillator has a single-well potential,
the bistability is a consequence of resonant modulation, and the
Hamiltonian $\hat g$ is independent of time. At the same time there
is a remote similarity between the oscillations of the tunneling
matrix element for period-2 states and for electron states in a
double-well potential in a quantizing magnetic field
\cite{Jain1988}. However, not only is the physics different, but our
approach is also different from that in Ref.~\onlinecite{Jain1988};
in particular, it makes it possible to find $\delta g$ analytically.
The approach can be also extended to a resonantly driven
Duffing oscillator, where the RWA Hamiltonian has a structure
similar to Eq.~(\ref{eq:g}) \cite{Dmitriev1986a,Dykman1979a}.

Tunnel splitting can be observed by preparing the system in one of
the period-2 states and by studying interstate oscillations, cf.
Refs.~\onlinecite{Hensinger2001,Steck2001}. This requires that the
tunneling rate $(\delta\omega/2\mu\lambda)|\delta g|$ exceed
$\omega_F/4{\cal Q}$, where ${\cal Q}$ is the oscillator quality
factor. The splitting sharply increases with increasing $\lambda$.
It will be shown separately that for comparatively large $\lambda$
(but still for $|\delta g|\ll g_q $) the RWA applies and relaxation
remains small provided $\delta g^2 \gg C_{\lambda}/{\cal Q}$ with
$C_{\lambda}\lesssim 1$. Our RWA numerical results indicate that
$\delta g$ still oscillates with $\mu$ for $\lambda =0.25-0.3$ and
is well described by Eq.~(\ref{eq:splitting_explicit}) for
$\mu\gtrsim 2\lambda$. The local peak of $|\delta g|$ for $\lambda
=0.3$ and the characteristic extremum of $d^2\delta g/d\mu^2$ for
$\lambda =0.25$ occur where $|\delta g|\approx 0.01$. Such $\delta
g$ may be large enough for detecting the effect in modulated
Josephson junctions where ${\cal Q}=2360$ has been reached in the
range of bistability \cite{Boaknin2007}.

In conclusion, we used the WKB approximation to study the wave
functions of the period-2 states of a parametrically modulated
oscillator. We showed that these wave functions can display spatial
oscillations in the classically forbidden region, in the rotating
frame. These oscillations lead to oscillations of the matrix element
of tunneling between the period-2 states with the varying frequency
of the modulating field.

We are grateful to V.N. Smelyanskiy and F. Wilhelm for stimulating
interactions. This research was supported in part by the NSF
through grant No. PHY-0555346.

%\bibliographystyle{apsrev}
%\bibliography{D:/Aaa/BibTex/md10}

\end{document}